\documentclass[structabstract]{aa}  
\usepackage[ansinew]{inputenc}
\usepackage{graphicx}
\usepackage{amsmath}
\usepackage{txfonts}
\usepackage{natbib}
\bibpunct{(}{)}{;}{a}{}{,}
\begin{document}
   \title{Mapping the prestellar core Ophiuchus D (L1696A) in ammonia\thanks{This publication is based on data acquired with the Very Large Array operated by The National Radio Astronomy Observatory (NRAO). NRAO is a facility of the National Science Foundation operated under cooperative agreement by Associated Universities, Inc.}}

   \author{J. Ruoskanen
          \inst{}
          ,
          J. Harju\inst{}
          ,
          M. Juvela\inst{}
          ,
          O. Miettinen\inst{}
          ,
          A. Liljestr\"{o}m\inst{}
		      ,
	        M. Väisälä\inst{}
          ,
          T. Lunttila\inst{}
          \and
          S. Kontinen
          }

   \institute{Department of Physics,
              P.O.Box 64, FI-00014 University of Helsinki, Finland\\
              \email{jukka.ruoskanen@helsinki.fi}
             }

   \date{Received; accepted}

 
  \abstract
   {The gas kinetic temperature in the centres of starless, high-density cores is predicted to fall as low as 5-6 K. The temperature gradient, which affects the dynamics and chemistry of these objects, should be discernible with radio interferometers reaching a spatial resolution of 1000 AU or better.}
   {The aim of this study was to determine the kinetic temperature distribution in the low-mass prestellar core Oph D where previous observations suggest a very low central temperature.}
   {The densest part of the Oph D core was mapped in the NH$_3$(1,1) and (2,2) inversion lines using the Very Large Array (VLA). The physical quantities were derived from the observed spectra by fitting the hyperfine structure of the lines, and subsequently the temperature distribution of Oph D was calculated using the standard rotational temperature techniques. A physical model of the cores was constructed, and the simulated spectra after radiative transfer calculations with a 3D Monte Carlo code were compared with the observed spectra. Temperature, density, and ammonia abundance of the core model were tuned until a satisfactory match with the observation was obtained.}
   {The high resolution of the interferometric data reveals that the southern part of Oph D comprises of two small cores in consistence with the 1.3mm dust continuum map of Motte et al. (1998, A\&A, 336, 150). The gas kinetic temperatures, as derived from ammonia towards the centres of the southern and northern core are 7.4 and 8.9 K, respectively. These values represent line-of-sight averages using the LTE assumption.  A model using modified Bonnor-Ebert spheres, in which the temperature decreases to 6.1 K and 8.9 K in the centres of southern and northern core, matched the observed values satisfactorily. The southern core, which has more steep temperature gradient, has central density of $n_{\mathrm{c}}=4\times 10^6$ cm$^{-3}$, and the data suggests depletion of ammonia within 700 AU from the centre. The northern core, which is almost isothermal, seems to be less dense. The radial velocity gradients in these cores are almost opposite in direction, which may be an indication that turbulent fragmentation has a role in the formation of these cores. The observed masses of the cores are only $\sim0.2$ M$_{\sun}$. Their potential collapse could lead to formation of brown dwarfs or low-mass stars.}
   {}

\titlerunning{Mapping the prestellar core Oph D in ammonia}
\authorrunning{Ruoskanen et al.}
   \maketitle
%

\section{Introduction}

Dense cores within molecular clouds are birth places of stars as some of them are gravitationally bound and will subsequently collapse and form a star. Observations of starless, prestellar cores promote our understanding of the initial stages of star formation. Molecular line observations provide information about e.g. gas density, cloud structure, and kinematics of the gas. Furthermore, they can be utilised to probe gas kinetic temperature. Radiative transfer calculations suggest, that the external heating by the interstellar radiation field is attenuated by dust in dense cores thus causing the gas temperature to fall towards the centre of these cores \citep[see,  e.g.][]{2001A&A...376..650Z,2001ApJ...557..193E,2005ApJ...635.1151K}. The temperature gradient is most pronounced within a radius of a few thousand AU from the core centre. Far infrared- and submillimeter dust continuum studies have shown dust temperature to decrease towards the centre of cores \citep[see,  e.g.][]{2002MNRAS.329..257W,2002A&A...382..583J,2003A&A...406L..59P,2004A&A...417..605P,2005ApJ...624..254S}. High angular resolution observations showing this behaviour for gas have not been extensively conducted \citep[one of the few is][]{2007A&A...470..221C}.

An isolated, roundish starless core with high central density ($n_\mathrm{c}>10^{5}$ cm$^{-3}$) is a preferred choice to look for the predicted temperature gradient. One nearby core that fulfills these criteria is Ophiuchus D (L1696A), north from the $\rho$-Ophiuchus molecular cloud. We adopt the new distance estimate of 120 pc to the $\rho$-Ophiuchus cloud for this core \citep{2008A&A...480..785L}. There is no embedded protostar in southern part of Oph D \citep{2007MNRAS.375..843K}. Our reference position RA = 16$^{\mathrm{h}}$28$^{\mathrm{m}}$28\fs 9, Dec = --24\degr 19 \arcmin 09\arcsec (J2000), coincides with the dust emission maximum of the 850 $\mu$m SCUBA map of \citet{2005MNRAS.360.1506K}, the dust absorption maximum of 7 $\mu$m ISOCAM map of \citet{2000A&A...361..555B}, and the N$_2$D$^+$ line emission peak according to the observation by \citet{2005ApJ...619..379C}. Direct evidence of very cold gas in molecular clouds was obtained by \citet{2008A&A...482..535H} when observing the \textit{ortho}-H$_2$D$^+$ line towards Oph D with APEX, using the same antenna pointing position as our reference point. They reported the kinetic temperature of the core derived from H$_2$D$^+$ linewidth to be 6.0$\pm$1.4 K.

Nitrogen bearing molecular species, such as NH$_3$, are found in high density cores ($n_\mathrm{c}>10^{5}$ cm$^{-3}$) where most carbon bearing species have already depleted from gas phase \citep[e.g.,][]{1998ApJ...507L.171W,2001ApJ...552..639A,2002ApJ...569..815T,2004A&A...416..603J}. In addition to tracing the densest gas NH$_3$ is also a good thermometer of the gas \citep{1979ApJ...234..912H}, and therefore appropriate choice for probing the temperature gradient.


\section{Observations and data reduction}

Maps of the \textit{para}-NH$_3$ (\textit{J,K}) = (1,1) and (2,2) transitions at frequencies 23\,694.4955 MHz and 23\,722.6336 MHz were made with the Very Large Array (VLA) in five observing slots, each six hours in duration, between 9th and 15th June 2008 in the array's DnC configuration. The synthesised beam of VLA in this hybrid compact configuration was 3\farcs72 $\times$ 2\farcs35 (position angle PA = 37\fdg2 east of north) with natural weighting applied. The FWHM of the primary beam is 2\arcmin\ at this frequency. In order to improve the signal-to-noise ratio of the (2,2) map we used a circular clean beam with FWHM of 10\arcsec\ to construct the clean image. For consistency we also cleaned the (1,1) map with the same beam to ensure identical resolution of the maps for the analysis. The transitions were observed simultaneously in 2IF mode with 128 channels and a bandwidth of 1.5625 MHz in both spectral windows (resulting in a channel width of 12.2 kHz or 0.154 km s$^{-1}$). The bandwidth was enough to include the main hyperfine group and the inner $F_1=1-2$ and $F_1=2-1$ satellite groups of the (1,1) line. 
   
Temporal variations in gain and phase were calibrated every 15 minutes using quasar 1626-298 ($F_\mathrm{1.3cm}$ = 2.57 Jy\footnote{\tiny{VLA calibration manual:\\{\tt http://www.vla.nrao.edu/astro/calib/manual/csource.html}}}). Bandpass- and absolute flux calibration were carried out in the beginning and the end of each 6 hour period using quasar 1331+305 ($F_\mathrm{1.3cm}$ = 2.59 Jy). The uncertainty of flux calibration was estimated to be of the order of 15\%. At the time of the observation some of the antennas of the array were replaced by new Expanded VLA antennas, for which reason the online Doppler tracking was not available. In order to avoid abrupt phase jumps the sky frequency was calculated by hand for each phase calibrator - source - phase calibrator triplet. 

The visibilities were calibrated and data reduction and imaging was carried out with Common Astronomy Software Applications (CASA)\footnote{\tiny{CASA is being developed in a collaboration led by NRAO as a main post-processing package for ALMA and Expanded VLA, see {\tt http://casa.nrao.edu}}}. The noise rms of the data cube cleaned with the 10\arcsec\ beam is 1 mJy/beam, which corresponds to about 22 mK. We were unable to estimate the missing flux since no single dish data was available (the maximum in the full resolution clean map is 0.11 K km s$^{-1}$ whereas the negative contours reach a level of -0.06 K km s$^{-1}$). This should not affect the uncertainties of physical quantities derived from line intensity ratios, e.g. optical thickness and kinetic temperature, unless there is differential flux loss between (1,1) and (2,2) spectral lines.

We fitted the hyperfine structure of NH$_3$(1,1) -line using our $\chi^2$ -minimisation routine to obtain estimates of line total optical thickness $\tau_{11}$, velocity $\varv_\mathrm{LSR}$, and linewidth (FWHM) $\Delta \varv$. The hyperfine structure of (2,2) -line could not be detected, so the main component was fitted with a single Gaussian line profile.
 

\section{Results}

\subsection{Integrated intensity maps}

\begin{figure}
\resizebox{\hsize}{!}{\includegraphics[bb=126 259 436 561,clip]{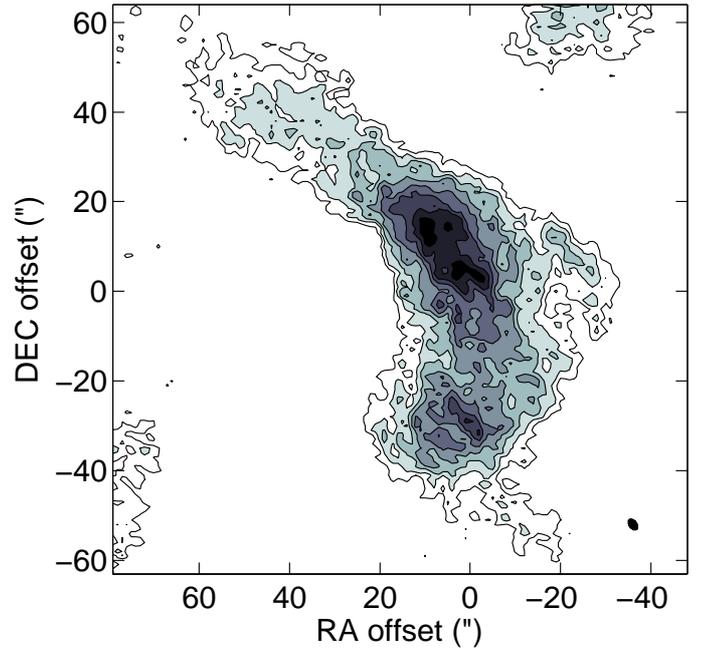}}
\caption{Integrated intensity map of Oph D in (1,1) transition of NH$_3$. The VLA synthesised beam, 3\farcs72 $\times$ 2\farcs35, is indicated in the bottom right corner. The contours are from 20\% to 90\% with steps of 10\% of the map peak value 0.11 K km s$^{-1}$. The reference position is RA = 16$^{\mathrm{h}}$28$^{\mathrm{m}}$28\fs 9, Dec = --24\degr 19 \arcmin 09\arcsec (J2000).}
\label{fig:fullreso}
\end{figure}

\begin{figure*}
\includegraphics[bb=-52 264 640 575,clip,width=18.3cm]{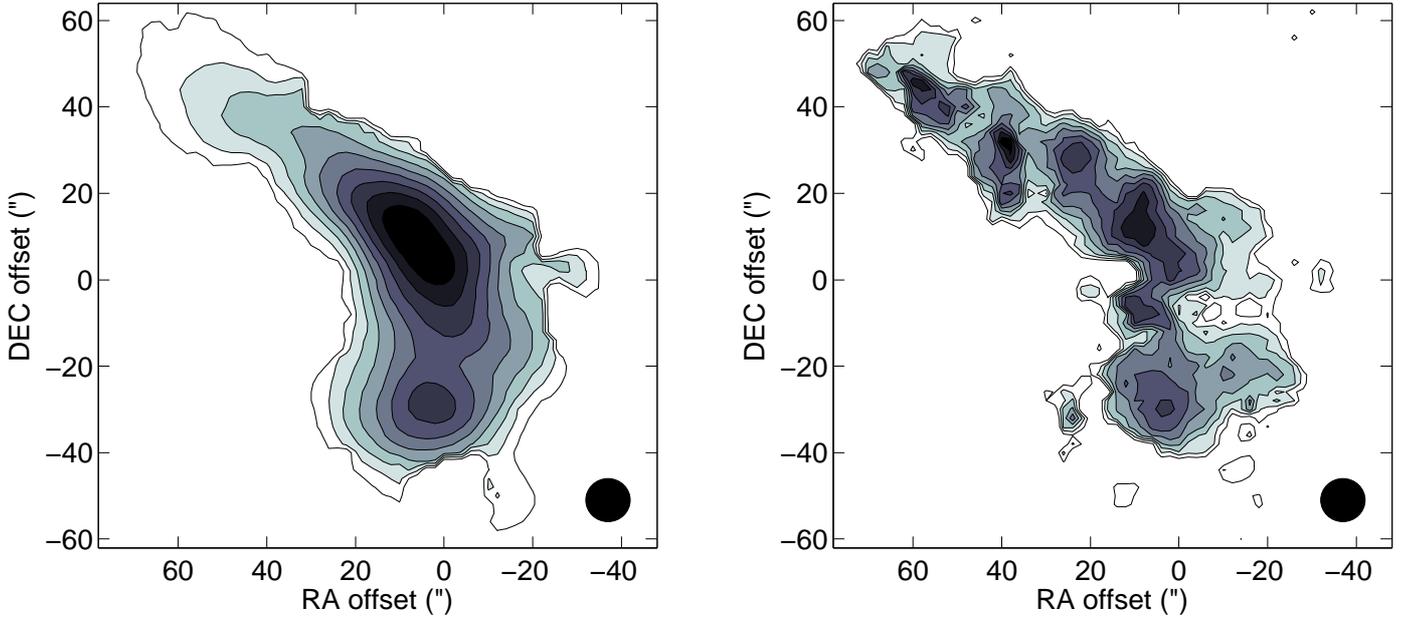}
\caption{Integrated intensity maps of Oph D in (1,1) (left) and (2,2) (right) transitions of NH$_3$ smoothed to 10\arcsec\ to improve the signal-to-noise ratio of (2,2) line. The 10\arcsec\ Gaussian clean beam is indicated in the bottom right corner. The contours are from 10\% to 90\% with steps of 10\% of the map peak values 3.5 K km s$^{-1}$ and 0.11 K km s$^{-1}$ for (1,1) and (2,2) lines, respectively.}
\label{fig:integrated}
\end{figure*}

The integrated intensity maps are shown in Figs. \ref{fig:fullreso} and \ref{fig:integrated}. The full angular resolution map of the ammonia (1,1) transition is shown in Fig. \ref{fig:fullreso}. The signal-to-noise ratio of the (2,2) transition was very small when cleaned to the full angular resolution, and therefore its map is not shown. In Fig. \ref{fig:integrated} the integrated intensity maps of (1,1) and (2,2) transitions smoothed to 10\arcsec\ are presented. The emission of (1,1) line was integrated over all available hyperfine components (main group and the inner satellite groups) and (2,2) line over the main component. The RA and DEC offsets in these maps are relative to our reference position.

The integrated intensity maps show two prominent local emission maxima within the mapped area, especially in NH$_3$(1,1) maps. Hereafter, we will refer to these as the northern and the southern peak. Such structure has not been seen in previous submillimeter or infrared observations nor in molecular line mappings of this core, with the exception of \citet{1998A&A...336..150M} who report two cores in the southern end of Oph D in their 1.3 mm continuum map (D-MM1 and D-MM2, see their Fig. 2). Typically the southern end of Oph D has been characterised as a single prestellar core. The structure will be discussed in more detail in section 4.

\begin{table}
\caption{Velocity gradients of Oph D measured from the NH$_3$(1,1) lines.}             
\label{table:velograd}      
\centering                          
\begin{tabular}{c l c c}        
\hline\hline                
Source\tablefootmark{a} & & $\mathcal{G}$ & $\Theta$\tablefootmark{b}\\    
 & & [km s$^{-1}$ pc$^{-1}$] & [$\degr$]\\    
\hline                       
   N & & 6.71$\pm$0.01  & 178.7$\pm$0.4\\      
   S & & 2.10$\pm$0.06  & 55.8$\pm$1.5\\
\hline                                   
\end{tabular}
\tablefoot{
\tablefoottext{a}{The northern peak is indicated with N, and southern with S.}
\tablefoottext{b}{Direction of increasing velocity measured west of north.}
}
\end{table}

\begin{figure}
\includegraphics[bb=104 229 474 594,clip,width=9.1cm]{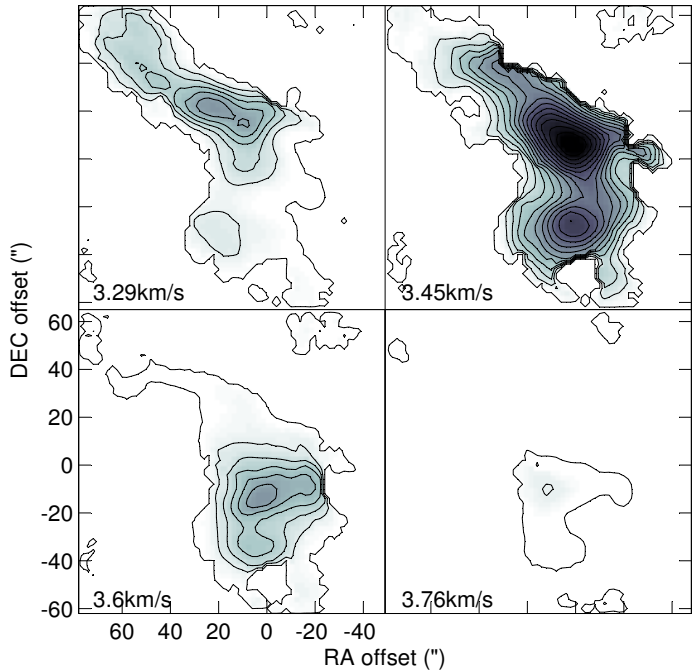}
\caption{Channel map of Oph D in the (1,1) transition of NH$_3$. Channel width is 0.154 km s$^{-1}$.}
\label{fig:chmaps}
\end{figure}

\subsection{Kinematics of the gas}

The kinematics of the gas was studied with the aid of Gaussian fits to the hyperfine structure of the NH$_3$(1,1) line which provide accurate estimates of radial velocity and line widths. The resulting $1\sigma$ uncertainties of these parameters are typically better than 0.01 km s$^{-1}$. The hyperfine fits were used to form another image cube with a single Gaussian line profile normalised with the integrated intensity of the full NH$_3$(1,1) line. This image cube describes the radial velocity distribution of the ammonia. The velocity channel map based on this image cube is shown in Fig. \ref{fig:chmaps}. The channel width corresponds to the original resolution of the VLA correlator. The map covers velocities from 3.29 to 3.76 km s$^{-1}$ in four channels. The emission of the northern peak moves from northeast towards southwest as the velocity increases, whereas the southern peak emission appears to move slowly westwards along a roughly east-west path.

\begin{figure}[t!]
\begin{center}
\includegraphics[bb=124 265 438 562,clip,width=9.1cm]{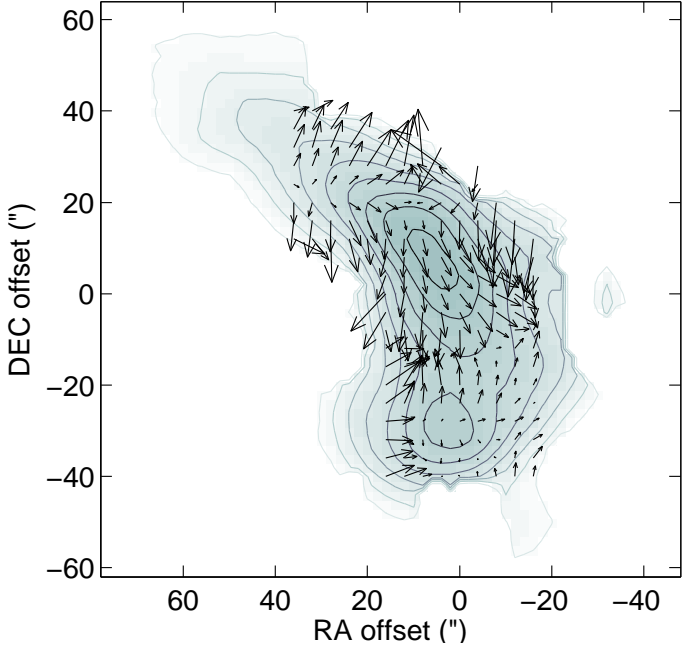}
\caption{Velocity gradient map of NH$_3$(1,1) transition of Oph D. The arrows depict the direction (of increasing radial velocity and the magnitude of the velocity gradients. An arrow with a length of 10\arcsec\ represents a gradient of 14.3 km s$^{-1}$ pc$^{-1}$. \label{fig:velo}}
\end{center}
\end{figure}

We created a local velocity gradient map for both transitions following the method presented in \citet{1993ApJ...406..528G}. Each velocity gradient is determined using a nine pixel grid centred at the pixel under consideration. Furthermore, we required that at least seven of the surrounding pixels have a usable velocity estimate available for the least squares fit. The results are presented in Fig \ref{fig:velo}, where the velocity gradient vectors of (1,1) transition are superimposed on the map of integrated intensity.

In the velocity gradient maps one can see two distinctly different velocity fields on top of the southern and the northern peaks, which seems to suggest dissimilar rotation of gas near these peaks. The effect can also be seen in the map of the (2,2) transition, although not as prominently. The velocity gradients in the vicinity of both peaks seem homogeneous enough to apply the solid-body rotation fit also in a larger scale. The overall total gradient is determined for the two locally uniform areas separately instead of giving one value to the whole clump. The magnitude and direction of total gradients for northern and southern peaks are given in Table \ref{table:velograd}. 

The magnitudes of the total velocity gradients, $\mathcal{G}=6.71$ and $\mathcal{G}=2.1$ km s$^{-1}$ pc$^{-1}$, of the northern and the southern peak are comparable to the values typically detected in other cores using nitrogen compound tracers  \citep{2002ApJ...572..238C,2002ApJ...565..331C,2004A&A...420..957C,2007A&A...470..221C}, but the agreement with previous observations of Oph D is poor \citep{1993ApJ...406..528G,2005ApJ...619..379C}. This is probably due to significant difference in the angular resolution of the data, being $\sim$ 80\arcsec\ \citep[see their Fig. 21]{1989ApJS...71...89B} for NH$_3$ data used in \citet{1993ApJ...406..528G} and 26\arcsec\ for the N$_2$H$^+$(1-0) data used in \citet{2005ApJ...619..379C}. With these beamwidths the velocity field is averaged considerably, and more importantly the reported magnitude of the total gradient of the whole core will not be a good figure of rotation for this object, since, as we have shown, there exists two distinct velocity regions with almost opposite gradient directions. 
\begin{table*}
\caption{NH$_3$(1,1) and (2,2) fits and derived parameters in the northern and the southern local column density maxima of Oph D using the 10\arcsec\ resolution map.}             
\label{table:fits}      
\centering                          
\begin{tabular}{l c c c c c c c c}        
\hline\hline                
Source / Line\tablefootmark{a} & $T_\mathrm{MB}$ & $\varv_\mathrm{LSR}$ & $\Delta\varv$ & $\tau_\mathrm{tot}$\tablefootmark{b} & $T_\mathrm{ex}$ & $T_{12}$ & $T_\mathrm{kin}$ & $N(\mbox{\tiny{NH$_3$}})$ \\    
 & [K] & [km s$^{-1}$] & [km s$^{-1}$] & & [K] & [K] & [K] & [10$^{14}$ cm$^{-2}$]\\    
\hline                       
   N, NH$_3$(1,1) & 3.52$\pm$0.06 & 3.39$\pm$0.001 & 0.18$\pm$0.003 & 10.37$\pm$0.39 & 6.05$\pm$0.50 & 8.59$\pm$0.37 & 8.87$\pm$0.56  & 9.60$\pm$2.97 \\
   N, NH$_3$(2,2) & 0.28$\pm$0.10 & 3.38$\pm$0.02  & 0.27$\pm$0.05  & & & & \\
   S, NH$_3$(1,1) & 3.33$\pm$0.12 & 3.47$\pm$0.002 & 0.16$\pm$0.005 & 7.16$\pm$0.52 & 5.70$\pm$0.50 & 7.32$\pm$0.59 & 7.43$\pm$1.12  & 9.44$\pm$2.75 \\
   S, NH$_3$(2,2) & 0.12$\pm$0.10 & 3.48$\pm$0.03  & 0.25$\pm$0.09  & & & & \\
\hline                                   
\end{tabular}
\tablefoot{
\tablefoottext{a}{The northern peak is indicated with N, and southern with S. The exact positions of the peaks are: RA = 16$^{\mathrm{h}}$28$^{\mathrm{m}}$29\fs 3, Dec = --24\degr 19 \arcmin 03\arcsec (J2000) for the northern and RA = 16$^{\mathrm{h}}$28$^{\mathrm{m}}$28\fs 8, Dec = --24\degr 19 \arcmin 23\arcsec for the southern peak.}
\tablefoottext{b}{The sum of peak optical thicknesses of all (1,1) hyperfine components. This is twice the 'main group opacity' used in many papers.}
}
\end{table*} 
\begin{figure}
\includegraphics[bb=41 219 532 602,clip,width=9.1cm]{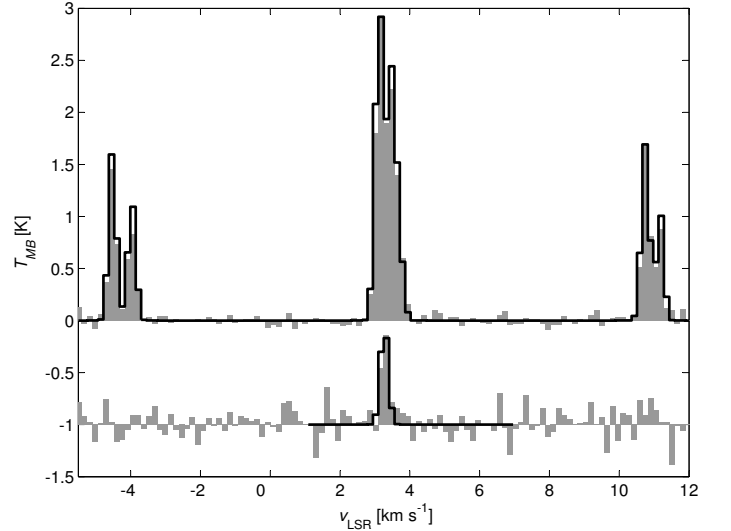}
\caption{Spectra (gray bars) and chi-square fits (black solid line) of NH$_3$(1,1) and (2,2) from the centre of the northern core (from the 10\arcsec\ resolution map). The (2,2) spectrum is offset by -1 K for clarity. The splitting of the (1,1) components is due to hyperfine structure, and is not to be taken as a sign of self-absorption.}
\label{fig:spec}
\end{figure}

\subsection{Ammonia column densities and the kinetic temperatures using the LTE assumption}

We fitted the hyperfine structure of NH$_3$(1,1) line using a $\chi^2$ routine, and the (2,2) line with a simple Gaussian profile. Fits were performed for pixels having signal larger than $3\sigma_\mathrm{noise}$ level. We made the usual assumption, that the different hyperfine states are populated according to LTE. We derived the excitation temperature $T_\mathrm{ex}$, rotational temperature $T_{12}$ and the total (ortho + para) ammonia column density $N(\mbox{NH$_3$})$ following the procedure described by \citet{1993A&AS...98...51H}. In the calculation of total ammonia column density we assumed the NH$_3$ ortho-to-para ratio to follow the LTE conditions. The kinetic temperature of the gas was calculated from the rotational temperature using the semi-empirically derived analytic expression from \citet{2004Ap&SS.292..347T}
\begin{equation}
T_\mathrm{kin}=\frac{T_{12}}{1-(T_{12}/42)\ln\left[1+1.1\exp(-16/T_{12})\right]},\label{eq:tafalla}
\end{equation}
which is based on Monte Carlo simulations using NH$_3$ - H$_2$ collision coefficients from \citet{1988MNRAS.235..229D} and is found to be suitable for dense cores having temperatures less than 20 K. According to \citet{2009MNRAS.399..425M}, who present new collisional rates for NH$_3$ - H$_2$, their new and Danby's coefficients agree within a factor of two. They also found, that non-LTE Monte Carlo radiative transfer simulations using their new and Danby's coefficients produce practically equal results for a parameter set typical of a dense and cold prestellar core. Thus, Eq. (\ref{eq:tafalla}) characterises the relation between rotational and kinetic temperature of the gas robustly. 

\begin{figure*}
\includegraphics[bb=40 309 577 527,clip,width=18.3cm]{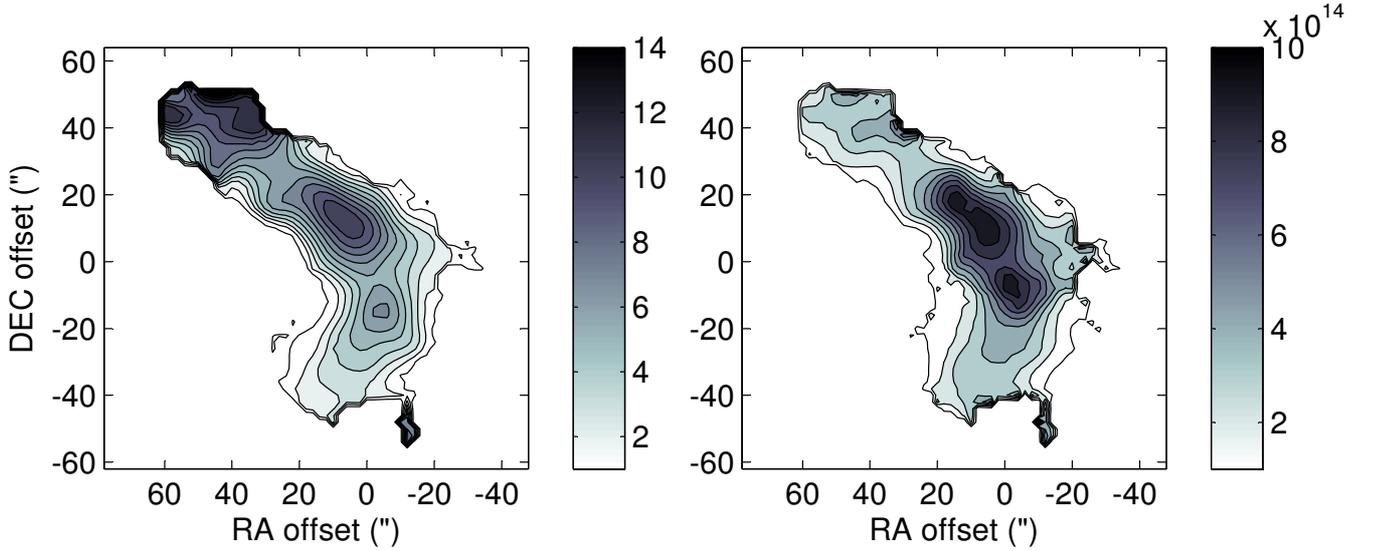}
\caption{The map of NH$_3$ optical thickness $\tau$ (left) and total column density $N$(NH$_3$) [cm$^{-2}$] (right) of Oph D.}
\label{fig:njatau}
\end{figure*}

The results of the hyperfine fits and the physical parameters derived for the northern peak (RA = 16$^{\mathrm{h}}$28$^{\mathrm{m}}$29\fs 3, Dec = --24\degr 19 \arcmin 03\arcsec) and the southern peak (RA = 16$^{\mathrm{h}}$28$^{\mathrm{m}}$28\fs 8, Dec = --24\degr 19 \arcmin 23\arcsec) peaks are presented in Table \ref{table:fits}. The maps of NH$_3$ optical thickness $\tau$ and column density $N(\mbox{NH$_3$})$ are shown in Fig. \ref{fig:njatau}.

\subsection{Mass and fractional abundance}

In the absence of continuum data of our own, we use the 1.3 mm data of \citet{1998A&A...336..150M} to estimate masses and fractional abundance of ammonia of our sources. Despite the difference of 5\arcsec\ between the coordinates of our core centres and the D-MM1 and D-MM2 reported by \citet{1998A&A...336..150M}, we consider them the same cores.

If we assume the dust is isothermal (having temperature $T_\mathrm{d}$) and the background emission has been subtracted from the observed intensity, the intensity of dust emission $I_\nu^\mathrm{dust}$ can be calculated using the equation of radiative transfer:
\begin{equation}
I_\nu^\mathrm{dust}=B_\nu(T_\mathrm{d})(1-e^{-\tau_\nu}).\label{eq:50}
\end{equation}
Thermal dust emission in the mm/submm region is optically thin, so Eq. (\ref{eq:50}) becomes $I_\nu^\mathrm{dust}\approx B_\nu(T_\mathrm{d})\tau_\nu$. The optical thickness $\tau_\nu$ is given by the product of total surface density $\Sigma$, the mass absorption coefficient per unit mass of dust $\kappa_\mathrm{d}$ and the dust-to-gas mass ratio $R_\mathrm{d}$. The surface density can be expressed with the aid of column density $\Sigma=N(\mathrm{H_2})\mu_\mathrm{H_2} m_\mathrm{H}$, where $\mu_\mathrm{H_2}$ is the mean molecular weight per H$_2$ molecule and $m_{\rm H}$ is the mass of the hydrogen atom. The column density as a function of dust intensity and temperature becomes
\begin{equation}
N(\mathrm{H_2})=\frac{I_\nu^\mathrm{dust}}{B_\nu(T_\mathrm{d})\mu_\mathrm{H_2} m_\mathrm{H}\kappa_\mathrm{d}R_\mathrm{d} }.\label{eq:51}
\end{equation}
The 1.3 mm flux density in a 15\arcsec\ beam reported by \citet{1998A&A...336..150M} is 45$\pm$5 and 45$\pm$15 mJy beam$^{-1}$ for southern and northern core, respectively. We assume that the gas and dust temperatures are equal, and use an average value of $T_\mathrm{d}=8$ K for the dust temperature for both cores. Adopting values of $\kappa_\mathrm{d}=0.5$ cm$^2$ g$^{-1}$ \citep{1994A&A...291..943O} and $R_\mathrm{d}=0.01$, the molecular hydrogen column density for the cores becomes
	\[N(\mathrm{H}_2)=5.3\pm2.0\times 10^{22}\ \mathrm{cm}^{-2} \mbox{  for the northern and}\]
\[	N(\mathrm{H}_2)=5.3\pm0.6\times 10^{22}\ \mathrm{cm}^{-2} \mbox{  for the southern core}.
\]

The fractional abundance of ammonia can be estimated using the calculated value of molecular hydrogen column density and our ammonia column density (Table \ref{table:fits}),
	\[X(\mathrm{NH}_3)= \frac{N(\mathrm{NH}_3)}{N(\mathrm{H}_2)}\approx 1.8\times 10^{-8}.
\]
The uncertainty of factor $\sim$2 is typically associated with the dust opacity coefficient. This affects directly the column density, mass, and abundance estimates.

\citet{1998A&A...336..150M} give mass estimates for both cores based on integrated flux densities. The reported masses of the cores are 0.15 M$_{\sun}$ and 0.16 M$_{\sun}$ for the southern and northern cores. Based on peak flux densities, the mass within the 15\arcsec\ beam can be estimated with
\begin{equation}
M=\frac{S_\nu d^2}{B_\nu(T_\mathrm{d})\kappa_\mathrm{d}R_\mathrm{d}},\label{eq:13mass}
\end{equation}
where $d=120$ pc is the distance to the source. This gives a mass estimate of $M=0.10$ M$_{\sun}$ for both cores. The 15\arcsec\ beam, however, encloses the cores only partially, and therefore the mass estimate is probably too small to describe the whole cores.

\section{Modelling}

\subsection{Description of the model: BE spheres}

The southern tip of Oph D cloud has in general been considered a single prestellar core. Our VLA data, however, indicates the existence of two cores instead of one. This hypothesis is backed up by the velocity gradient map (Fig. \ref{fig:velo}) as well as by the two separate local maxima of optical thickness and column density seen in Fig. \ref{fig:njatau}. The velocity gradients seem to imply, that the direction and magnitude of rotation of the gas within the area of these two hypothetical cores is markedly different from each other. The secondary peaks of optical thickness and column density south from the most dominant peak also point towards the presence of two cores. \citet{2005A&A...434..167S} modelled Oph D cloud with a complex 3D substructure based on ISOCAM 7 and 15 $\mu$m maps and a 1.3 mm map from IRAM 30 m telescope. In their model the southern part of the cloud also contains two dominant density concentrations. Also \citet{1998A&A...336..150M} report two cores (D-MM1 and D-MM2 in their Fig. 2 and Table 2) in the southern part of Oph D.

We attempt to model the southern part of Oph D with two cores, each in hydrostatic equilibrium and with a radial temperature gradient, the so called modified Bonnor-Ebert spheres \citep{2001ApJ...557..193E,2001A&A...376..650Z} with radii 2200 AU (northern core) and 2100 AU (southern core). The central positions of the core models were chosen to coincide with the local kinetic temperature minima derived from the observation. The calculation of the gas density and temperature profiles for both core models was done iteratively until satisfactory convergence was achieved. The initial core model was calculated by \citet[see Sec. 2.1. therein]{2010A&A...509A..98S}, who starting from an isothermal sphere derived the dust density and temperature profile for a core with a radius of 2400 AU. The southern core density is high enough ($n_\mathrm{c}>10^6$ cm$^{-3}$) to assume the dust and gas temperatures to be equal \citep{1983ApJ...265..223B}, but for the northern core ($n_\mathrm{c}\sim 10^5$ cm$^{-3}$) the situation may not be as clear. Despite this uncertainty we used Sipilä's dust temperature and density profile, scaled to the radii of our core models, as a first estimate of our cores.

In order to derive the gas density profile the following modification to the original substitutions introduced by \citet{1956MNRAS.116..351B} was made to include the non-isothermal temperature distribution $T(r)$ of the core in the calculations \citep{2010A&A...509A..98S,2011arXiv1108.5275S}:
\begin{equation}
\rho(r)=\lambda\left(\frac{T_c}{T(r)}\right)\mbox{e}^{-\psi},\label{eq:bonnor}
\end{equation}
\begin{equation}
r=\xi\sqrt{\frac{kT_\mathrm{c}}{4\pi Gm\lambda}},\label{eq:bonnor2}
\end{equation}
where $T_c$ and $\lambda$ are the temperature and density at the centre. The resulting equation of hydrostatic equilibrium was solved numerically using boundary conditions presented in \citet{2010A&A...509A..98S}, and the obtained density and temperature profiles were fed to the radiative transfer Monte Carlo code \citep{1997A&A...322..943J} to simulate the spectra. We use molecular data provided by the Leiden Atomic and Molecular Database \citep{2005A&A...432..369S}. The method described in \citet{2004ApJ...613..355K} is used for handling overlapping hyperfine emission lines.
\begin{figure}
\includegraphics[bb=55 211 555 606,clip,width=9.1cm]{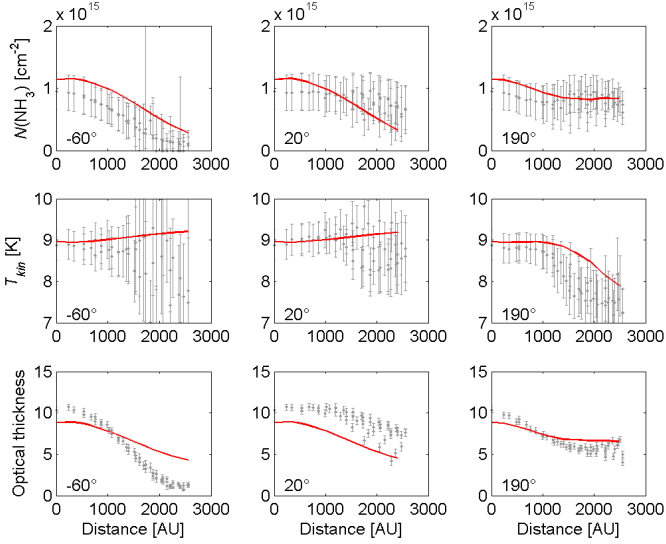}
\caption{Line-of-sight average profiles of column density, kinetic temperature and optical thickness ($\tau_\mathrm{tot}$) towards three radial directions (E of N) from northern core centre outwards. The values derived from the observation are marked with small gray circles and from the model with a solid line.}
\label{fig:modeln}
\end{figure}
For the simulation we constructed a 3D model which consists of these two cores with
a projected separation of 2600 AU in the plane of the sky. In addition to the density and temperature, the fractional abundance of ammonia $X(\mbox{NH$_3$})$ in our model can be adjusted, as well as the turbulent velocity within the core. The non-thermal velocity dispersion component $\Delta \varv_\mathrm{NT}$ is calculated from the observed linewidths $\Delta\varv_\mathrm{obs}$ with an equation:
\begin{equation}
\Delta\varv_\mathrm{NT}=\sqrt{(\Delta\varv_\mathrm{obs})^2-8\ln2\frac{k_\mathrm{B} T_\mathrm{kin}}{\mu_{\mbox{\tiny{NH}}_3}m_\mathrm{H}}}.\label{eq:velodisp}
\end{equation}
The prevailing non-thermal velocity dispersion in the northern and southern cores is between 40-50 m~s$^{-1}$ (the isothermal sound speed for a core with average gas kinetic temperature of 8 K is $c_s\sim$ 170 m~s$^{-1}$). The turbulent velocities in our simulation model are therefore also kept at a low value.
\begin{figure}
\includegraphics[bb=55 211 555 606,clip,width=9.1cm]{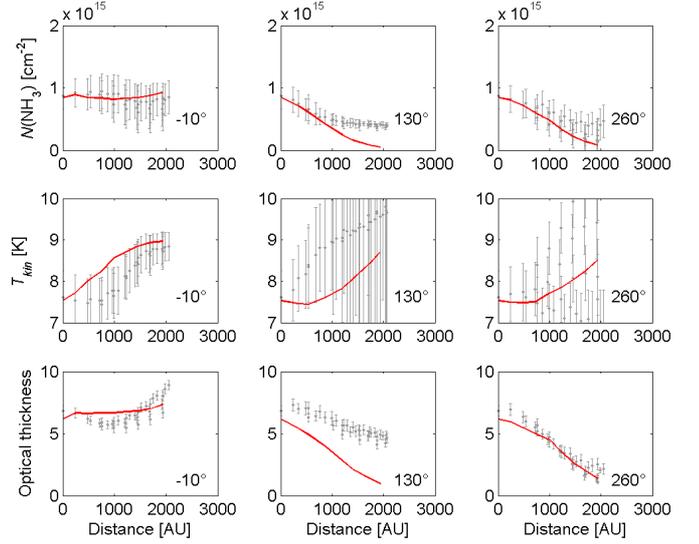}
\caption{Line-of-sight average profiles of column density, kinetic temperature and optical thickness ($\tau_\mathrm{tot}$) towards three radial directions (E of N) from southern core centre outwards. The values derived from the observation are marked with small gray circles and from the model with a solid line.}
\label{fig:models}
\end{figure}
\subsubsection{Comparison between the model and observation}

In dense prestellar cores the fractional abundance of NH$_3$ may increase towards the core centre \citep[see e.g.][and references therein]{2004A&A...415.1065H}. In more evolved cores, however, when the central density increases above $n_\mathrm{c}\sim10^6$ cm$^{-3}$, also ammonia starts to vanish from the gas phase \citep{2005ApJ...620..330A,2006A&A...456..215F,2010A&A...524A..91M}. A variety of fractional abundance values was tested in core models, using the estimate $X(\mathrm{NH}_3)\approx 1.5\times 10^{-8}$ discussed in section 3.4 as a starting point. In each simulation the abundance radial profile shape followed the current density profile shape, except in the case of depletion which approximated as a steep descending slope.

The best-fit parameters of the northern core turned it into a nearly isothermal sphere; the temperature at the centre is 8.8 K and increases only 0.4 K towards the outer edge. The optical thickness and column density could not be fitted simultaneously to match the observed values at the centre of the cores, and a relative error of $\sim$ 15\% remains. The central density of the model is $n_\mathrm{c}=2\times10^5$ cm$^{-3}$ and the fractional abundance gradually increases towards the centre peaking at the value $X(\mbox{NH${_3}$})=2\times10^{-8}$. Mass of the northern core model is only 0.045 M$_{\sun}$, and could not be increased while maintaining a reasonable agreement with the observation.

In order to model the low temperatures observed in the southern core the model central density had to be increased to $4\times10^6$ cm$^{-3}$, similar to the results of \citet{2008A&A...482..535H}. The best-fit temperature profile has a temperature decrease from 10 K to 6.1 K from the surface to the centre. The fractional abundance of ammonia was then used to control the optical thickness and column density. It was quickly noticed that a swift decrease in abundance within the innermost 700 AU was required to obtain a match with the observed values. The abundance increases to a value of $X(\mbox{NH${_3}$})=2.5\times10^{-9}$ before the depletion starts to take effect. The mass of the southern core model is 0.16 M$_{\sun}$. 

\begin{figure}[t!]
\begin{center}
\includegraphics[bb=48 232 568 595,clip,width=7.1cm]{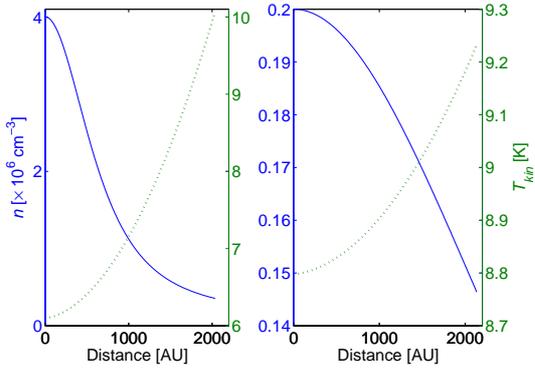}
\caption{The density (solid line) and temperature (dash line) profiles of the southern (left) and northern core (right) BE model. \label{fig:params}}
\end{center}
\end{figure}

The structure of the southern part of Oph D is certainly more complex than what can be modelled using two spherical cores in hydrostatic equilibrium. Nevertheless, a relatively close agreement with the observation was achieved with this simple model. The spectra produced by the radiative transfer calculations were analysed in the same way as the observed spectra. The results are presented in Figs. \ref{fig:modeln} and \ref{fig:models}. In both figures the column density, kinetic temperature, and optical thickness are plotted in three radial directions from the core centre towards the edge. Each subplot contains the observed values in a sector $\pm20\degr$ from the mean direction plotted with small gray circles together with their 1$\sigma$ uncertainties. The corresponding simulation result is shown with a solid line. The temperature and density of northern and southern core models are presented in Fig. \ref{fig:params}.

The 3D model was also used to reproduce the H$_2$D$^+$ line observed by \citet{2008A&A...482..535H} with APEX. The fractional abundance of H$_2$D$^+$ was taken from \citet{2010A&A...509A..98S} and applied for both cores. Otherwise the model was exactly the same as in NH$_3$ simulations. A single pointing spectrum at our and Harju's reference position using APEX antenna beam FWHM of 17\arcsec\ was simulated, and a linewidth of $\Delta\varv$ = 0.35 km s$^{-1}$ was obtained. This is exactly the same linewidth produced by a single hydrostatic core model of \citet{2008A&A...482..535H} and \citet{2010A&A...509A..98S} in their respective simulations of this observation. Although the observed narrow linewidth $\Delta\varv$ = 0.26 km s$^{-1}$ could not be reproduced with our two core model, we can at least see that the simulation is in line with the previous single core models, and did not introduce further deviation from the observed value.
\subsection{Description of an alternative model: spheroid cores}

The modified Bonnor-Ebert sphere models reproduce the observed spectra reasonably well, and is also an apparent choice for the roundish southern core. The shape of the northern core, however, appears more elongated in the VLA map. An oblate spheroid, whose axis of rotation is on a plane perpendicular to our line of sight, would describe the shape and velocity field of the northern core perhaps better than a sphere. To test what effect the rotation of the cloud has on its shape, we run simulations with the Pencil Code\footnote{\tiny{{\tt http://code.google.com/p/pencil-code/}}} using a 3D-grid. Pencil Code is a modular solver of magnetohydrodynamic equations using a high order finite difference scheme. The solution method enables a straightforward combination of various ''physics modules'', like self-gravity, and various equations of state to the general equations describing the time-evolution of a hydrodynamic flow. As an initial condition, we used an isothermal Bonnor-Ebert sphere, where an initial rotation rate corresponded to the angular velocity of the northern core $\Omega = 2.2 \times 10^{-13}$ rad s$^{-1}$. No extra rotational energy was put into the system during the time development of the simulation, and magnetic fields were neglected. The rotation rate $t_{\mathrm{0}}\Omega=0.27$, where the dynamical timescale for BE sphere is $t_0=(4\pi G \rho_{\mathrm{c}})^{-1/2}$ \citep{2004MNRAS.355..248B}, is quite fast, but still not enough to play alone a significant role in supporting the core against gravitational collapse. The dynamics of the model over time were observed, and with this angular velocity, according to calculations, the core seems to settle in an spheroidical equilibrium state.

The experiment indicated that the Bonnor-Ebert sphere transforms into a spheroid (with an axial ratio of the projected ellipse less than 2:1). The convergence in the simulation is slow, however, and still after 1 Myr there are oscillations. At this point the central density has decreased to one fifth of the original value, and the density profile is somewhat constant along the longer axis until a rather sudden drop occurs at the edge of the core. In the direction of the axis of rotation the shape of the original hydrostatic equilibrium density distribution remains about the same, just rescaled. 

In this alternative model a slightly spheroidical shape was adopted for the southern core as well. The density and temperature profiles of the spheroids are, as was indicated by the simulation conducted with the rotating sphere, radially scaled versions of those of modified Bonnor-Ebert spheres. In this model the central densities were $n_\mathrm{c}=1.0\times10^6$ cm$^{-3}$ and $n_\mathrm{c}=1.4\times10^6$ cm$^{-3}$ for the northern and southern cores. The fractional abundance in both cores has a decrease in the centre, i.e. ammonia is depleted. The northern core abundance reaches a peak value of $X(\mbox{NH${_3}$})=1.1\times10^{-8}$ 1000 AU from the centre and the southern core abundance $X(\mbox{NH${_3}$})=7\times10^{-9}$ 700 AU from the centre.

The best-fit temperature profile of the northern core is steeper than in the previously discussed Bonnor-Ebert sphere model. The model gas temperature decreases to 7.4 K in the centre of the core, and rises to 10.2 K at the surface. The southern core temperature profile has a minimum at the centre of 6.4 K and rises to 8.0 K towards the surface. Masses of the models are 0.11 M$_{\sun}$ and 0.07 M$_{\sun}$ for northern and southern core, respectively.

\subsection{Comparison between the two models}

The simulation results of the two spheroid model show that the temperature profile of the northern core is not in good agreement with the observed values. The column density and optical thickness, on the other hand, are slightly better matched than with the modified Bonnor-Ebert sphere model. The central depletion of ammonia and higher gas density seem to provide more promising match than the less dense core with incresing abundance towards the centre.

Both models of the southern core have problems with optical thickness and temperature profiles. The modified Bonnor-Ebert sphere model seems to do better with the temperature whereas with the optical thickness profile both models fail to explain the observed values consistently. 

We estimated the goodness of fit by calculating the reduced $\chi^2$ value between the model profile and the data:
\begin{equation}
\chi^2=\frac{1}{D}\sum_{i=1}^n \frac{(O_i-E_i)^2}{\sigma_i^2}, \label{eq:chis}
\end{equation}
where $D$ is the number of degrees of freedom, $O$ the observed value, $E$ the modelled value and $\sigma^2$ the variance of observation. The $\chi^2$ -values are presented in Table \ref{table:chis1}. The comparison between the two models reveals no clear winner; both models have strong and weak properties. For the southern core, however, the use of a modified Bonnor-Ebert sphere leads to better results. 

\begin{table}
\caption{Comparison of $\chi^2$ -values of fitted models.}             
\label{table:chis1}      
\centering                          
\begin{tabular}{l c c c}        
\hline\hline                
 & $N(\mathrm{NH}_3)$ & $T_\mathrm{kin}$ & $\tau_{11}$\\    
\hline                       
   Northern, model 1 \tablefootmark{a} & 1.38 & 1.12 & 1.87 \\
   Northern, model 2 \tablefootmark{b} & 1.28 & 1.70 & 1.61 \\
   Southern, model 1 \tablefootmark{a} & 0.91 & 1.34 & 1.58 \\
   Southern, model 2 \tablefootmark{b} & 1.44 & 1.58 & 2.03 \\
\hline                                   
\end{tabular}
\tablefoot{
\tablefoottext{a}{The model with modified Bonnor-Ebert spheres.}
\tablefoottext{b}{The model with spheroids.}
}
\end{table} 

It should be noted, however, that models with only one core, be it a spheroid or modified Bonnor-Ebert sphere, did not produce acceptable match with the observed values. Also, if we dismiss the temperature gradient of the southern core and make it isothermal, the temperature estimated from the simulated ammonia spectra is severely different from the observation. This indicates that a temperature gradient is necessary in the physical model of the southern core. The northern core is almost isothermal, and its best-fit radial temperature profile has only 0.4 K difference between centre and surface.
\begin{table*}
\caption{Physical parameters of the two cores in the southern part of Oph D.}             
\label{table:physparams}      
\centering                          
\begin{tabular}{l c c c c c c c c c}        
\hline\hline                
Core & $M_\mathrm{model}$\tablefootmark{a} & $M_\mathrm{obs}$\tablefootmark{b} & $M_\mathrm{vir}$ & $T_\mathrm{kin}$\tablefootmark{c} & $n_{\mathrm{c}}$\tablefootmark{d} & $E_\mathrm{therm}/\vert E_\mathrm{grav}\vert$\tablefootmark{e} & $E_\mathrm{rot}/\vert E_\mathrm{grav}\vert$ & $E_{\mathrm{turb}}/\vert E_\mathrm{grav}\vert$  & $P_\mathrm{ext}/k_B$\tablefootmark{f}\\    
 & [M$_{\sun}$] & [M$_{\sun}$] &[M$_{\sun}$] & [K] & [10$^{6}$ cm$^{-3}$] & & & & [K cm$^{-3}$]\\    
\hline                       
   Northern & 0.045 & 0.21 & 0.29 & 8.8 & 0.2 & 3.89 (0.497) & 0.077 & 0.045& 2.0$\times10^6$\\
   Southern & 0.16 & 0.17 & 0.21 & 6.1 & 4.0 & 0.75 (0.413) & 1.23$\times10^{-3}$ & 1.0$\times10^{-2}$ & 9.1$\times10^6$\\
\hline                                   
\end{tabular}
\tablefoot{
\tablefoottext{a}{The mass of the modified Bonnor-Ebert sphere used to model the core.}
\tablefoottext{b}{The mass of the core derived from the observation with the aid of the model. The projection area of the core and the column density of ammonia within that area were taken from the observation. In order to calculate the mass, a value for the fractional abundance of ammonia was taken from the model (modified BE sphere). The average values of abundance over the projected area of the model were estimated to be $X$(NH$_3$) = $1.8\times 10^{-8}$ for the northern and $X$(NH$_3$) = $5\times 10^{-9}$ for the southern core.}
\tablefoottext{c}{The kinetic temperature at the centre of the core model.}
\tablefoottext{d}{The number density at the centre of the core model.}
\tablefoottext{e}{Ratio of thermal energy to gravitational potential energy using the core model mass $M_\mathrm{model}$, and in parenthesis using $M_\mathrm{obs}$.}
\tablefoottext{f}{The external pressure, when using the core mass derived from observation ($M_\mathrm{obs}=0.21$ M$_{\sun}$), is $P_\mathrm{ext}/k_B=9.9\times10^6$ K cm$^{-3}$.}
}
\end{table*} 


\section{Discussion}

As was seen in the previous section, our distinctly different models both produce a reasonable match with the observation, but none of the two can be clearly pointed superior. The model with two hydrostatic cores, modified Bonnor-Ebert spheres, is, however, on more solid ground theoretically and we will commit ourselves to that model in the following discussion.

A Bonnor-Ebert sphere satisfies the following virial equilibrium equation
\begin{equation}
2T+U=3VP_\mathrm{ext}, \label{eq:virial}
\end{equation}
where $P_\mathrm{ext}$ is the external pressure, and $T$ and $U$ the kinetic and gravitational potential energies. We neglect the effects of possible magnetic field, and calculate the kinetic energy
\begin{equation}
T=E_{\mathrm{therm}}+E_{\mathrm{rot}}+E_{\mathrm{turb}}=\frac{3}{2}M\frac{k_{\mathrm{B}}T_{\rm kin}}{\mu m_{\mathrm{H}}} + \frac{1}{2}I\Omega^2 + \frac{3}{2}M\frac{(\Delta \varv_{\mathrm{NT}})^2}{8\ln 2} \label{eq:T}
\end{equation}
and gravitational potential energy
\begin{equation}
U=E_{\mathrm{grav}}=\frac{-GM^2}{R}f(\xi_0)=\frac{-GM^2}{R}\left[\frac{3}{\xi_0\frac{d\psi}{d\xi_0}}-\frac{e^{-\psi}}{\left(\frac{d\psi}{d\xi_0}\right)^2}\right] \label{eq:U}
\end{equation}
of the cores to examine their relations. From the relation (\ref{eq:virial}) we can estimate the external pressure assuming that the cores are Bonnor-Ebert spheres. The centrally peaked density profile was taken into account individually for both cores in the numerical calculation of moment of inertia, $I$, in the equation of $E_{\mathrm{rot}}$. The function $f(\xi_0)$ takes the density profile into account in gravitational potential energy for an isothermal core. We will use it although our cores have a temperature gradient (the slight difference between the isothermal and non-isothermal BE spheres is illustrated in Appendix B of \citet{2011arXiv1108.5275S}). The value of $f(\xi_0)$ for a homogeneous density core is 0.6. The southern one of our core models is slightly supercritical $\xi_0=6.75$ ($f(\xi_0)=0.74$) and the northern one has a markedly lower value $\xi_0=1.49$ ($f(\xi_0)=0.61$). The energy ratios are presented in Table \ref{table:physparams}.

The ratio of thermal energy to gravitational energy of the northern core is 3.89. This would suggest, that the external pressure inhibits the core from dispersing. On the other hand, when the core mass is estimated using data partly from the observation and partly from the model the energy ratio is $\sim$ 0.5. The core mass in this case is estimated using the column densities within the area of projected ellipse in the interferometric map. When the average fractional abundance is estimated from the model Bonnor-Ebert sphere ($X$(NH$_3$) = $1.8\times 10^{-8}$ for the northern and $X$(NH$_3$) = $5\times 10^{-9}$ for the southern core), the total mass of the core can be calculated. The difference between this mass and the model mass for the northern core suggests that the spheroidal ellipsoid into which an initially rotating BE sphere evolves to, might be a more suitable core model. However, the mass of the spheroid model was also very low, only 0.11 M$_{\sun}$. 

The modified BE sphere mass of the southern core is almost equal to the mass derived from the observations as described above. The southern core seems to be gravitationally bound, which can be seen from the ratio of kinetic and gravitational energy in Table \ref{table:physparams}. The mass of the core is very close to the estimation given by \citet{1998A&A...336..150M}, 0.15 M$_{\sun}$. Also the peak flux density of Motte's 1.3 mm continuum data was equal for both cores. Due to their vicinity the external pressure for both cores should be roughly the same. The values are close when the northern core mass is the one derived from the observation. Using the model mass the difference is significant (see Table \ref{table:physparams}). These facts further support the conclusion, that the northern core model is not correct, despite the fact that the match with the observed quantities is satisfactory.

\section{Summary and conclusions}

We have presented an interferometric observation of the southern part of Ophiuchus D prestellar core. We mapped the core with VLA using the NH$_3$(1,1) and (2,2) inversion lines. The synthetic beam of VLA allowed spatial resolution in the order of 400 AU, but we were forced to decrease it to about 1000 AU in order to improve the (2,2) line signal-to-noise ratio. From the observed spectra we derived the gas kinetic temperature distribution within the mapped area. The total column density of ammonia, the optical thickness of both lines, and the radial velocity distributions were also obtained.

Our observation shows that the southern part of Oph D comprises of two potentially prestellar cores. In the southern core the line-of-sight average gas temperature derived from the observed ammonia spectra drops from the outer edge towards the centre, reaching a minimum of $T_{\mathrm{kin}}=7.4$ K. Towards the centre of the northern core the line-of-sight average gas temperature from ammonia is $T_{\mathrm{kin}}=8.9$ K. The northern core is almost isothermal with only 0.4 K increase in temperature from centre to the edge. The observed masses of the cores, 0.21 M$_{\sun}$ for the northern and 0.17 M$_{\sun}$ for the southern core, are quite small. Their potential collapse could lead to the formation of brown dwarfs or low mass stars. The two cores were previously detected by \citet{1998A&A...336..150M}.

Our observation shows that there exists two rather homogeneous velocity gradient fields within the mapped area. They coincide well with the positions of the northern and southern cores. If we interpret the velocity gradients as rotational motion, we see that the two cores are rotating in almost opposite directions. The angular velocity of the northern core is $\Omega=2.2\times 10^{-13}$ rad s$^{-1}$ and the southern $\Omega=6.9\times 10^{-12}$ rad s$^{-1}$. \citet{2005A&A...434..167S} proposed that Oph D would have been formed by turbulent fragmentation. The directions and magnitudes of the gas velocity field derived from our observation support this view.

A modified Bonnor-Ebert sphere approximates the southern core reasonably well. The density and temperature profiles are reproduced satisfactorily when the central number density of the model is $n_{\mathrm{c}}=4\times 10^6$ cm$^{-3}$ and the model temperature drops from 10 K at the surface to 6.1 K in the centre. Heavy ammonia depletion within 700 AU from the centre is required to succesfully match the southern core model with the observation.

The northern core is more challenging to model, and we did not find a single good solution. Two different parameter value combinations succeed in producing an acceptable match with the observation. The other one is a modified Bonnor-Ebert sphere with central density of $n_{\mathrm{c}}=2\times 10^5$ cm$^{-3}$ and having ammonia abundance increase inwards, reaching $X(\mbox{NH${_3}$})=2\times10^{-8}$ at the centre. An almost equally good match with the observation can be achieved with a more dense spheroidical core with $n_{\mathrm{c}}=1\times 10^6$ cm$^{-3}$ and heavy ammonia depletion within the innermost 1000 AU. Gas kinetic temperatures at the centres of the two core models are 8.8 K and 7.4 K, respectively.

The location of the cold spot in the southern part of Oph D according to \citet{2008A&A...482..535H} is almost in between our two cores. They observed the ground state ($1_{10} \rightarrow 1_{11}$) rotational transition of \textit{ortho}-H$_2$D$^+$, which may peak at a different position than ammonia owing to chemistry. Nevertheless, the previous H$_2$D$^+$ result is surprising as one might expect to find the coldest gas to concentrate in the core centres. 

The minimum line-of-sight average gas kinetic temperature derived from the interferometric NH$_3$ observations is $T_\mathrm{kin}=7.4 \pm 1.1 $ K. The difference between this result and the temperature reported by \citet{2008A&A...482..535H} ($T_\mathrm{kin}=6.0 \pm 1.4 $ K) is not statistically significant, and can furthermore be understood by considering the following points. First of all, H$_2$D$^+$ probably traces denser, and thus colder gas than ammonia. Furthermore, our temperature represents the line-of-sight average, which inevitably increases the temperature estimate as ammonia thrives also in warmer envelope of the core. The fact that we could not utilise the full resolution interferometric map, but were forced to use a map cleaned with a 10\arcsec\ restoring beam in our analysis, also influenced our results.

The cores seem to be gravitationally bound. The observed angular velocities of both cores, especially the northern one, are significant. If there is enough time for the cores to evolve, the rotation may lead to a formation of a disk. The southern core simulation results suggest that the core may be eventually collapsing, as the dimensionless radius of the Bonnor-Ebert sphere is $\xi=6.75$, slightly above the critical value. There is no observational data to aid in the discussion regarding the role of magnetic field in the support of these cores. Some mechanisms, like a slight increase in the external pressure and, in the case of magnetic support, ambipolar diffusion is needed to trigger the collapse process.

\begin{acknowledgements}
The study has been funded by the Academy of Finland through grants 132291 and 127015. We thank all the VLA staff, especially Gustav van Moorsel, for their hospitality and support during our stay at the VLA. We are grateful to the anonymous referee for helpful comments. We also thank the editor, Malcolm Walmsley, for his useful comments. This research has made use of NASA's Astrophysics Data System and the NASA/IPAC Infrared Science Archive, which is operated by the JPL, California Institute of Technology, under contract with the NASA.
\end{acknowledgements}
\bibliographystyle{aa} 
\small
\bibliography{ophD} 

\begin{thebibliography}{44}
\expandafter\ifx\csname natexlab\endcsname\relax\def\natexlab#1{#1}\fi

\bibitem[{{Aikawa} {et~al.}(2005){Aikawa}, {Herbst}, {Roberts}, \&
  {Caselli}}]{2005ApJ...620..330A}
{Aikawa}, Y., {Herbst}, E., {Roberts}, H., \& {Caselli}, P. 2005, \apj, 620,
  330

\bibitem[{{Aikawa} {et~al.}(2001){Aikawa}, {Ohashi}, {Inutsuka}, {Herbst}, \&
  {Takakuwa}}]{2001ApJ...552..639A}
{Aikawa}, Y., {Ohashi}, N., {Inutsuka}, S., {Herbst}, E., \& {Takakuwa}, S.
  2001, \apj, 552, 639

\bibitem[{{Bacmann} {et~al.}(2000){Bacmann}, {Andr{\'e}}, {Puget}, {Abergel},
  {Bontemps}, \& {Ward-Thompson}}]{2000A&A...361..555B}
{Bacmann}, A., {Andr{\'e}}, P., {Puget}, J., {et~al.} 2000, \aap, 361, 555

\bibitem[{{Banerjee} {et~al.}(2004){Banerjee}, {Pudritz}, \&
  {Holmes}}]{2004MNRAS.355..248B}
{Banerjee}, R., {Pudritz}, R.~E., \& {Holmes}, L. 2004, \mnras, 355, 248

\bibitem[{{Benson} \& {Myers}(1989)}]{1989ApJS...71...89B}
{Benson}, P.~J. \& {Myers}, P.~C. 1989, \apjs, 71, 89

\bibitem[{{Bonnor}(1956)}]{1956MNRAS.116..351B}
{Bonnor}, W.~B. 1956, \mnras, 116, 351

\bibitem[{{Burke} \& {Hollenbach}(1983)}]{1983ApJ...265..223B}
{Burke}, J.~R. \& {Hollenbach}, D.~J. 1983, \apj, 265, 223

\bibitem[{{Caselli} {et~al.}(2002{\natexlab{a}}){Caselli}, {Benson}, {Myers},
  \& {Tafalla}}]{2002ApJ...572..238C}
{Caselli}, P., {Benson}, P.~J., {Myers}, P.~C., \& {Tafalla}, M.
  2002{\natexlab{a}}, \apj, 572, 238

\bibitem[{{Caselli} {et~al.}(2002{\natexlab{b}}){Caselli}, {Walmsley},
  {Zucconi}, {Tafalla}, {Dore}, \& {Myers}}]{2002ApJ...565..331C}
{Caselli}, P., {Walmsley}, C.~M., {Zucconi}, A., {et~al.} 2002{\natexlab{b}},
  \apj, 565, 331

\bibitem[{{Crapsi} {et~al.}(2005){Crapsi}, {Caselli}, {Walmsley}, {Myers},
  {Tafalla}, {Lee}, \& {Bourke}}]{2005ApJ...619..379C}
{Crapsi}, A., {Caselli}, P., {Walmsley}, C.~M., {et~al.} 2005, \apj, 619, 379

\bibitem[{{Crapsi} {et~al.}(2004){Crapsi}, {Caselli}, {Walmsley}, {Tafalla},
  {Lee}, {Bourke}, \& {Myers}}]{2004A&A...420..957C}
{Crapsi}, A., {Caselli}, P., {Walmsley}, C.~M., {et~al.} 2004, \aap, 420, 957

\bibitem[{{Crapsi} {et~al.}(2007){Crapsi}, {Caselli}, {Walmsley}, \&
  {Tafalla}}]{2007A&A...470..221C}
{Crapsi}, A., {Caselli}, P., {Walmsley}, M.~C., \& {Tafalla}, M. 2007, \aap,
  470, 221

\bibitem[{{Danby} {et~al.}(1988){Danby}, {Flower}, {Valiron}, {Schilke}, \&
  {Walmsley}}]{1988MNRAS.235..229D}
{Danby}, G., {Flower}, D.~R., {Valiron}, P., {Schilke}, P., \& {Walmsley},
  C.~M. 1988, \mnras, 235, 229

\bibitem[{{Evans} {et~al.}(2001){Evans}, {Rawlings}, {Shirley}, \&
  {Mundy}}]{2001ApJ...557..193E}
{Evans}, II, N.~J., {Rawlings}, J.~M.~C., {Shirley}, Y.~L., \& {Mundy}, L.~G.
  2001, \apj, 557, 193

\bibitem[{{Flower} {et~al.}(2006){Flower}, {Pineau Des For{\^e}ts}, \&
  {Walmsley}}]{2006A&A...456..215F}
{Flower}, D.~R., {Pineau Des For{\^e}ts}, G., \& {Walmsley}, C.~M. 2006, \aap,
  456, 215

\bibitem[{{Goodman} {et~al.}(1993){Goodman}, {Benson}, {Fuller}, \&
  {Myers}}]{1993ApJ...406..528G}
{Goodman}, A.~A., {Benson}, P.~J., {Fuller}, G.~A., \& {Myers}, P.~C. 1993,
  \apj, 406, 528

\bibitem[{{Harju} {et~al.}(2008){Harju}, {Juvela}, {Schlemmer}, {Haikala},
  {Lehtinen}, \& {Mattila}}]{2008A&A...482..535H}
{Harju}, J., {Juvela}, M., {Schlemmer}, S., {et~al.} 2008, \aap, 482, 535

\bibitem[{{Harju} {et~al.}(1993){Harju}, {Walmsley}, \&
  {Wouterloot}}]{1993A&AS...98...51H}
{Harju}, J., {Walmsley}, C.~M., \& {Wouterloot}, J.~G.~A. 1993, \aaps, 98, 51

\bibitem[{{Ho} {et~al.}(1979){Ho}, {Barrett}, {Myers}, {Matsakis}, {Chui},
  {Townes}, {Cheung}, \& {Yngvesson}}]{1979ApJ...234..912H}
{Ho}, P.~T.~P., {Barrett}, A.~H., {Myers}, P.~C., {et~al.} 1979, \apj, 234, 912

\bibitem[{{Hotzel} {et~al.}(2004){Hotzel}, {Harju}, \&
  {Walmsley}}]{2004A&A...415.1065H}
{Hotzel}, S., {Harju}, J., \& {Walmsley}, C.~M. 2004, \aap, 415, 1065

\bibitem[{{J{\o}rgensen} {et~al.}(2004){J{\o}rgensen}, {Sch{\"o}ier}, \& {van
  Dishoeck}}]{2004A&A...416..603J}
{J{\o}rgensen}, J.~K., {Sch{\"o}ier}, F.~L., \& {van Dishoeck}, E.~F. 2004,
  \aap, 416, 603

\bibitem[{{Juvela}(1997)}]{1997A&A...322..943J}
{Juvela}, M. 1997, \aap, 322, 943

\bibitem[{{Juvela} {et~al.}(2002){Juvela}, {Mattila}, {Lehtinen}, {Lemke},
  {Laureijs}, \& {Prusti}}]{2002A&A...382..583J}
{Juvela}, M., {Mattila}, K., {Lehtinen}, K., {et~al.} 2002, \aap, 382, 583

\bibitem[{{Keto} \& {Field}(2005)}]{2005ApJ...635.1151K}
{Keto}, E. \& {Field}, G. 2005, \apj, 635, 1151

\bibitem[{{Keto} {et~al.}(2004){Keto}, {Rybicki}, {Bergin}, \&
  {Plume}}]{2004ApJ...613..355K}
{Keto}, E., {Rybicki}, G.~B., {Bergin}, E.~A., \& {Plume}, R. 2004, \apj, 613,
  355

\bibitem[{{Kirk} {et~al.}(2005){Kirk}, {Ward-Thompson}, \&
  {Andr{\'e}}}]{2005MNRAS.360.1506K}
{Kirk}, J.~M., {Ward-Thompson}, D., \& {Andr{\'e}}, P. 2005, \mnras, 360, 1506

\bibitem[{{Kirk} {et~al.}(2007){Kirk}, {Ward-Thompson}, \&
  {Andr{\'e}}}]{2007MNRAS.375..843K}
{Kirk}, J.~M., {Ward-Thompson}, D., \& {Andr{\'e}}, P. 2007, \mnras, 375, 843

\bibitem[{{Lombardi} {et~al.}(2008){Lombardi}, {Lada}, \&
  {Alves}}]{2008A&A...480..785L}
{Lombardi}, M., {Lada}, C.~J., \& {Alves}, J. 2008, \aap, 480, 785

\bibitem[{{Maret} {et~al.}(2009){Maret}, {Faure}, {Scifoni}, \&
  {Wiesenfeld}}]{2009MNRAS.399..425M}
{Maret}, S., {Faure}, A., {Scifoni}, E., \& {Wiesenfeld}, L. 2009, \mnras, 399,
  425

\bibitem[{{Miettinen} {et~al.}(2010){Miettinen}, {Harju}, {Haikala}, \&
  {Juvela}}]{2010A&A...524A..91M}
{Miettinen}, O., {Harju}, J., {Haikala}, L.~K., \& {Juvela}, M. 2010, \aap,
  524, A91+

\bibitem[{{Motte} {et~al.}(1998){Motte}, {Andre}, \&
  {Neri}}]{1998A&A...336..150M}
{Motte}, F., {Andre}, P., \& {Neri}, R. 1998, \aap, 336, 150

\bibitem[{{Ossenkopf} \& {Henning}(1994)}]{1994A&A...291..943O}
{Ossenkopf}, V. \& {Henning}, T. 1994, \aap, 291, 943

\bibitem[{{Pagani} {et~al.}(2004){Pagani}, {Bacmann}, {Motte}, {Cambr{\'e}sy},
  {Fich}, {Lagache}, {Miville-Desch{\^e}nes}, {Pardo}, \&
  {Apponi}}]{2004A&A...417..605P}
{Pagani}, L., {Bacmann}, A., {Motte}, F., {et~al.} 2004, \aap, 417, 605

\bibitem[{{Pagani} {et~al.}(2003){Pagani}, {Lagache}, {Bacmann}, {Motte},
  {Cambr{\'e}sy}, {Fich}, {Teyssier}, {Miville-Desch{\^e}nes}, {Pardo},
  {Apponi}, \& {Stepnik}}]{2003A&A...406L..59P}
{Pagani}, L., {Lagache}, G., {Bacmann}, A., {et~al.} 2003, \aap, 406, L59

\bibitem[{{Schnee} \& {Goodman}(2005)}]{2005ApJ...624..254S}
{Schnee}, S. \& {Goodman}, A. 2005, \apj, 624, 254

\bibitem[{{Sch{\"o}ier} {et~al.}(2005){Sch{\"o}ier}, {van der Tak}, {van
  Dishoeck}, \& {Black}}]{2005A&A...432..369S}
{Sch{\"o}ier}, F.~L., {van der Tak}, F.~F.~S., {van Dishoeck}, E.~F., \&
  {Black}, J.~H. 2005, \aap, 432, 369

\bibitem[{{Sipil{\"a}} {et~al.}(2011){Sipil{\"a}}, {Harju}, \&
  {Juvela}}]{2011arXiv1108.5275S}
{Sipil{\"a}}, O., {Harju}, J., \& {Juvela}, M. 2011, \aap, in press

\bibitem[{{Sipil{\"a}} {et~al.}(2010){Sipil{\"a}}, {Hugo}, {Harju}, {Asvany},
  {Juvela}, \& {Schlemmer}}]{2010A&A...509A..98S}
{Sipil{\"a}}, O., {Hugo}, E., {Harju}, J., {et~al.} 2010, \aap, 509, A98+

\bibitem[{{Steinacker} {et~al.}(2005){Steinacker}, {Bacmann}, {Henning},
  {Klessen}, \& {Stickel}}]{2005A&A...434..167S}
{Steinacker}, J., {Bacmann}, A., {Henning}, T., {Klessen}, R., \& {Stickel}, M.
  2005, \aap, 434, 167

\bibitem[{{Tafalla} {et~al.}(2004){Tafalla}, {Myers}, {Caselli}, \&
  {Walmsley}}]{2004Ap&SS.292..347T}
{Tafalla}, M., {Myers}, P.~C., {Caselli}, P., \& {Walmsley}, C.~M. 2004, \apss,
  292, 347

\bibitem[{{Tafalla} {et~al.}(2002){Tafalla}, {Myers}, {Caselli}, {Walmsley}, \&
  {Comito}}]{2002ApJ...569..815T}
{Tafalla}, M., {Myers}, P.~C., {Caselli}, P., {Walmsley}, C.~M., \& {Comito},
  C. 2002, \apj, 569, 815

\bibitem[{{Ward-Thompson} {et~al.}(2002){Ward-Thompson}, {Andr{\'e}}, \&
  {Kirk}}]{2002MNRAS.329..257W}
{Ward-Thompson}, D., {Andr{\'e}}, P., \& {Kirk}, J.~M. 2002, \mnras, 329, 257

\bibitem[{{Willacy} {et~al.}(1998){Willacy}, {Langer}, \&
  {Velusamy}}]{1998ApJ...507L.171W}
{Willacy}, K., {Langer}, W.~D., \& {Velusamy}, T. 1998, \apjl, 507, L171

\bibitem[{{Zucconi} {et~al.}(2001){Zucconi}, {Walmsley}, \&
  {Galli}}]{2001A&A...376..650Z}
{Zucconi}, A., {Walmsley}, C.~M., \& {Galli}, D. 2001, \aap, 376, 650

\end{thebibliography}

\end{document}